\documentclass[structabstract]{aa}  
\usepackage{graphicx}
\usepackage{txfonts}
\usepackage{natbib}
\begin{document}
\newcommand{\isor}{$^6\mathrm{Li}/^7\mathrm{Li}$}
\newcommand{\litl}{ion{Li}{i} 670.78}

   \title{$^6${Li}/$^7${Li} estimates for metal-poor stars}

   \author{A.~E. Garc{\'\i}a P\'erez
          \inst{1}
          \and
          {W. Aoki}       
          \inst{2,3}
          \and
          {S. Inoue}
          \inst{4}
          \and
          {S.~G. Ryan}
          \inst{1}
          \and
          {T. ~K. Suzuki}
          \inst{5}
          \and
          {M. Chiba}
          \inst{6}
          }

   \institute{Centre for Astrophysics Research, STRI and School of Physics, Astronomy and Mathematics, University of Hertfordshire,
              College Lane, Hatfield AL10 9AB, United Kingdom\\
              \email{a.e.garcia-perez@herts.ac.uk, s.g.ryan@herts.ac.uk}
         \and
             National Astronomical Observatory, Mitaka, Tokyo 181-8588, Japan\\
             \email{aoki.wako@nao.ac.jp}
           \and
            Department of Astronomical Science, Graduate University of Advanced
            Studies, Mitaka, Tokyo 181-8588, Japan
         \and
             Department of Physics, Kyoto University, Oiwake-cho, Kitashirakawa, Sakyo-ku, Kyoto 606-8502, Japan
             \email{inoue@tap.scphys.kyoto-u.ac.jp}
         \and
             Department of Earth Science and Astronomy, Graduate School of Arts and Science, University of Tokyo, 3-8-1 Komaba, Meguro, Tokyo 153-8902, Japan
              \email{stakeru@ea.c.u-tokyo.ac.jp}
         \and
             Astronomical Institute, Tohoku University, Sendai 980-8578, Japan
             \email{chiba@astr.tohoku.ac.jp}
             }
   \date{; }

% \abstract{}{}{}{}{} 
% 5 {} token are mandatory
 
  \abstract
   {The presence of the lithium-6 isotope in some metal-poor stars is a matter of surprise because of the high values observed. Non-standard models of Big Bang nucleosynthesis and pre-Galactic cosmic ray fusion and spallation have been proposed to explain these values. However, the observations of this light isotope are challenging which may make some detections disputable.}
   {The goal was to determine {\isor} for a sample of metal-poor stars; three of them have been previously studied and the remaining two are new for this type of study. The purpose was to increase, if possible, the number of lithium-6 detections and to confirm previously published results.}
   {Spectra of the resonance doublet line of neutral lithium \ion{Li}{i} 670.78~nm  were taken with the High Dispersion Spectrograph at the Subaru 8.2~m-telescope for a sample of five metal-poor stars (${-3.12\le\mathrm{[Fe/H]}\le-2.19}$). The contribution of lithium-6 to the total observed line profile was estimated from the 1D-LTE analysis of the line asymmetry.}
   {Observed asymmetries could be reproduced assuming isotopic abundance ratios $^6\mathrm{Li}/^7\mathrm{Li}$ of the order of: $0.004$ for \object{BD $+26^\circ\,3578$}, $\sim 0.010$ for \object{BD $+02^\circ\,3375$} and \object{G 64-37}, $0.025$ for \object{BD $+20^\circ\,3603$} and $0.047$ for \object{BD $-04^\circ\,3208$}. We found that these results were very sensitive to several of the assumptions made in the analysis, in particular, the treatment of the residual structure in the analysed spectra. Our final estimates for the errors are respectively $\Delta${\isor}\ $= \pm 0.028, 0.029, 0.039, 0.025$ and $0.039$. }
   {The {\isor} ratios for the sample are comparable to or even lower than these error values, so that detections of lithium-6 can not safely be claimed despite of the high resolving power ($R \sim 95\,000$) and $S/N$ (400-600).}


   \maketitle
%
%________________________________________________________________

\section{Introduction}

The abundances of light elements in metal-poor stars are a good indicator of what happened early in the Universe and the Galaxy \citep{Alpha48}. The origins of the two stable isotopes of the light element lithium, $^6\mathrm{Li}$  and $^7\mathrm{Li}$, are still puzzling. Their derived abundances in metal-poor stars are difficult to reconcile simultaneously with standard predictions. Different production sites in the early Galaxy have been discussed: Big Bang nucleosynthesis \citep{Wagoner67, Nollett97, Coc04,Jedamzik08}, cosmic ray spallation and/or $\alpha$-fusion \citep{Reeves70,Vangioni00} and neutrino spallation (see e.g. \citealt{Vangioni96}). The most widely discussed picture is that lithium-6 has a cosmic-ray-induced origin (principally $\alpha$-fusion) \citep[and references therein]{Vangioni00, Aoki04, Asplund06, Rollinde06} and lithium-7 a predominantly primordial one \citep{Spite82}. 

The observed abundances of $^6\mathrm{Li}$ are used to discriminate between different production sites: pre-Galactic (e.g. \citealt{Suzuki02, Rollinde06, Rollinde08}), the Galactic interstellar medium (e.g. {\citealt{Meneguzzi71, Fields99}}), superbubbles (e.g. \citealt{Parizot99,Vangioni00}), fast supernova eject (e.g. \citealt{Fields96, Nakamura06}), in-situ stellar flares \citep{Tatischeff07}, etc. The problem with this isotope is that detections for metal-poor stars are very challenging and very rare. \citet{Smith93} reported the first probable detection of $^6\mathrm{Li}$ in HD 84937 ($^6\mathrm{Li}/^7\mathrm{Li} = 0.05\pm0.02$) which was later confirmed by other works \citep{Hobbs94, Smith98, Cayrel99}. A few years after this first detection, a detection for another star, \object{BD $+26^\circ\,3578$ ($^6\mathrm{Li}/^7\mathrm{Li} = 0.05\pm0.03$}, \citealt{Smith98}), was reported. Possible detections for two other stars were provided by \cite{Cayrel99} and \cite{Nissen00}: BD$+42^\circ\,2667$ ($^6\mathrm{Li}/^7\mathrm{Li} = 0.10$) and {G\,271-162} ($^6\mathrm{Li}/^7\mathrm{Li} = 0.02\pm0.01$) respectively. There are results published for other metal-poor stars also, however the values are mostly upper limits. Our group determined $^6\mathrm{Li}/^7\mathrm{Li}< 0.018$ for the metal-poor star HD140283 \citep{Aoki04}; this star may have partly depleted its initial content \citep{Cayrel99}. The work of \citet{Asplund06} has the highest number of published detections in metal-poor stars: nine. The detections for these stars appear to lie in a plateau that, if confirmed, would be difficult to explain within the standard framework of light element production. One would expect an evolution of the lithium-6 abundances with metallicity because of the contribution of Galactic cosmic ray spallation. Analyses of $^6$Li for very metal-poor stars ($\mathrm{[Fe/H]}<-2.5$) are still very limited, so new samples of a few stars have a large impact. 

A possible solution to the apparent existence of a plateau and also to the high abundance of that plateau could lie in a  pre-Galactic origin of the cosmic rays \citep{Suzuki02,Rollinde06} (see, however \citealt{Evoli08}). Non-standard models of Big Bang nucleosynthesis (BBN), generally involving particles of uncertain character, are also considered as a possible solution. The importance of establishing whether or not $^6$Li has been reliably detected in metal-poor stars is thus readily apparent: the previously published detections have been a significant motivator of investigations of the role of supersymmetry in modifying nucleosynthesis during the Big Bang. The detection (or otherwise) of $^6$Li could therefore potentially provide crucial, rare observational constraints on the nature of certain hypothesised supersymmetric particles and dark matter candidates, some of which cannot be studied experimentally with current or near-term experimental facilities (e.g. \citealt{Bailly08}). 

Although detections for  $^7\mathrm{Li}$ are more common than for $^6\mathrm{Li}$, the origin of  the heavier isotope is also quite intriguing. \citet{Spite82} observed essentially the same lithium content, the so called Spite value, in a sample of metal-poor dwarf stars with effective temperatures $5800\le T_\mathrm{eff}\le 6250$, despite their metallicities ranging from $-2.2\le\mathrm{[Fe/H]}\le-1.1$. A plausible interpretation for the plateau is that it is close to the primordial value.  $^7\mathrm{Li}$ is the only light isotope besides H, D, $^4$He and $^3$He that, according to standard BBN models, is predicted to be produced in significant amounts in the Big Bang. Predictions for the primordial lithium content based on standard Big Bang nucleosynthesis (SBBN) in combination with the baryon density inferred from the {\it{Wilkinson Microwave Anisotropy Probe Mission}} (WMAP) give a value two or three times higher than the observed plateau value, $A\mathrm{(Li)}\footnote{$A(\mathrm{Li}) = \log{N_{\mathrm{Li}}/N_\mathrm{H}}$+12} = 2.65$ vs $2.1$-$2.3$\,dex \citep[and references therein]{Coc04, Asplund06, Hosford09, Aoki09}. 

Lithium is destroyed by nuclear reactions at high temperatures. In the presence of mixing, material from the stellar atmospheres can reach depths of high temperatures where lithium can burn or can mix with material that has burnt it. According to some stellar models, little or no lithium-7 destruction is expected for stars in the Spite plateau (e.g. \citealt{Deliyannis90} and references therein) so that these stars should retain their initial content. However atomic diffusion is expected to act in warm main-sequence stars and alter the initial photospheric chemical composition (e.g. \citealt{Richard02,Richard05}). Stellar models including diffusion and/or stellar rotation have been suggested to reconcile the observations of metal-poor stars with the SBBN predictions, e.g \citet{Pinsonneault99} and \citet{Richard05}. Observers have searched for signatures of mixing by measuring the lithium-7 abundance dispersion and by studying the behaviour of the abundances with stellar temperatures and metallicity. There is not consensus yet whether the Spite plateau is a plateau or not and whether lithium abundances are dispersed or not (e.g. \citealt{Bonif97, Ryan99, Ryan00, Asplund06, Bonifacio07, Hosford09, Aoki09}). Lithium-6 is however more fragile than lithium-7 (burning at temperatures $T \sim 1.6\times10^6$\,K versus $T \sim 2.0\times10^6$\,K), so an observable content of the lighter isotope sets constraints on the degree of depletion of the other one. Stellar models without diffusion or rotational mixing predict part destruction of lithium-6 in the pre-main-sequence phase for most of the stars with published lithium-6 detections \citep{Proffitt89,Deliyannis90, Asplund06}. According to \citet{Asplund06} neither the rotational mixing models of \citet{Pinsonneault99} nor the diffusion models (including turbulent diffusion) from \citet{Richard02,Richard05} seem to give a solution to both lithium problems simultaneously. A high depletion of lithium-6 would be difficult to reconcile with observations of the other stable isotope because a very high initial content of lithium-6 would generally imply a significant lithium-7 production by cosmic ray nucleosynthesis. Nevertheless, recent results for extremely metal-poor stars \citep{Bonifacio07} show a correlation between lithium-7 abundance and metallicity which may have an interpretation in terms of Galactic evolution by cosmic ray spallation.

Given the importance of the published lithium-6 detections for the understanding of Big Bang and cosmic ray nucleosynthesis, it is paramount to confirm the detections and, if possible, to add more. This is particularly important because of the very weak signal presented by $^6\mathrm{Li}$: a small depression of the red wing of the $^7\mathrm{Li}$ doublet, at a level comparable to possible sources of systematic error. We present here $^6\mathrm{Li}/^7\mathrm{Li}$ estimates based on observations taken with the Subaru 8.2m-telescope for five metal-poor stars, of which three are in common with previous studies. The observations are described in \S~\ref{obs}, while the stellar parameter determinations and the spectral synthesis are in \S~\ref{secstepar}. In \S~\ref{secbroad} we present the stellar and instrumental broadening calculations. In \S~\ref{seciso} we explain the procedure followed to get our best isotopic abundance ratio estimates, and we present the results and their sensitivity to different uncertainties, while in \S~\ref{secdis} we discuss the results in terms of the observational quality and the sensitivity to the spectral analysis. We finish with conclusions in \S~\ref{seccon}.

%__________________________________________________________________

\section{\label{obs} Observations}

The $^6\mathrm{Li}/^7\mathrm{Li}$ estimates presented here are based on the spectral analysis of the observed \ion{Li}{i} resonance doublet line at 670.78\,nm. Spectra were taken for eight metal-poor stars with the High Dispersion Spectrograph (HDS, \citealt{Noguchi02}) mounted on the Subaru 8.2\,m-telescope on the nights of May 17-19, 2005 (UT). Results for five stars are presented here; the observations of three other stars were not completed in this run to
achieve the data quality required for this study. Values for exposure times are listed in Table\,\ref{stepar}, together with the $V$ magnitude, date of the observations, exposure time, the central wavelength of the instrumental setting employed and the measured heliocentric radial velocity of the star. Total exposures were split into several shorter ones for cosmic ray rejection. Three of the stars were observed in two consecutive nights. Observations were taken using a $2048\times4100$ EEV CCD ( $13.5\,\mu$m pixel size), a slit width of 0.4\,arcsec (0.20\,mm) and two standard instrumental settings: one with central wavelength at 608.11\,nm (StdYd) and the other at 620.78\,nm (StdYb) (0.002\,nm/pixel dispersion). Special care was taken for the calibration; a high number of flat-field frames ($\sim20$) were acquired and ThAr lamp spectra were taken at several times. Fast rotator stars were also observed, in case their spectra were required for flat-fielding correction.  

The observational data were reduced under {\sc{iraf}} using the echelle package and following the standard procedure: bias subtraction, flat-fielding correction, spectra extraction, wavelength calibration, normalisation and combination of individual spectra (spectral orders were not merged). For the wavelength calibration, an average of  20-40 Th lines per spectral order were identified. The wavelengths of these lines were used to convert pixel values into wavelengths. The wavelength solutions are based on a fifth order polynomial fit to the transformation. Typical values for the rms of these fits are 0.06\,pm. If necessary, the fits were interpolated in time before being applied to the observed spectra. The normalisation consisted of dividing the spectra by their cubic spline fit with a wavelength scale of 0.3\,nm. Spectra were co-added and weighted according to their exposure times. 

The co-added spectra are of very high quality. The signal-to-noise ratio ($S/N$) per pixel was estimated from the measurements of noise on both sides of \ion{Li}{i} 670.78\,nm line in a window of 0.07\,nm, values are in the range 417-611. The resolving power ($R = \lambda/\Delta\lambda_\mathrm{FWHM}$) was estimated from the full width at half maximum (FWHM) of a Gaussian fit to a set of ThAr lines lying in the spectral order that contains the lithium resonance doublet line; results are in the range $R \sim $90\,000-100\,000.
  
\section{\label{secstepar} Stellar parameters, model atmospheres and spectral synthesis}

The observed spectra were analysed using spectral synthesis. The first step in the analysis was to get an estimate of the fundamental parameters describing the stellar atmospheres: effective temperature $T_\mathrm{eff}$, the surface gravity  $\log g$, the metallicity $\mathrm{[Fe}/\mathrm{H}]$ and the microturbulence $\xi_\mathrm{t}$. The values listed in Table\,\ref{stepar} were taken from \cite{Nissen02, Nissen04}, with the exception of the metallicity. The choice of the stellar parameters is not very important for the accuracy of the $^6\mathrm{Li}/^7\mathrm{Li}$ final results (e.g. \citealt{Smith93,Asplund06}), however we have tried to have the maximum consistency. Although for some stars there are more recent values than the values assumed here, we have chosen those which were determined using the same indicators and in a similar way: $(b-y)$ and $(V-K)$ colors for $T_\mathrm{eff}$, and Hipparcos parallaxes for $\log{g}$. The uncertainty in the measured parallax of \object{G 64-37} is very high so an estimate coming from the \citet{Schuster04} calibration of the Str\"omgren indices based on Hipparcos parallaxes was used instead. Our metallicity estimates come from  the measured equivalent widths of a set of \ion{Fe}{ii} lines: $\lambda\,614.92$, $\lambda\,624.76$, $\lambda\,641.69$, $\lambda\,643.27$, and $\lambda\,645.64$\,nm. Iron abundances were derived and taken as a metallicity estimate. The solar values were determined from the measured equivalent widths of these lines as in \cite{Garcia06} [but now using new 3D solar C and O abundance values]. For \object{G 64-37}, the \ion{Fe}{ii} lines above were not detected, but one at 523.4\,nm was, giving $\mathrm{[Fe/H]}=-2.97$. As our metallicity calculation for this star would be based on only one line, we preferred to take the value from \cite{Nissen04} that is determined from several lines: $\mathrm{[Fe/H]} = -3.12$.

Equipped with the stellar parameters (see Table\,\ref{stepar}), model atmospheres were computed using the {\sc{marcs}} code version in \citet{Asplund97}. The code solves the equations of hydrostatic equilibrium and energy equilibrium assuming that the atmospheres have either plane-parallel or spherical symmetry (1D); we assumed plane-parallel geometry in our computations. The radiation and the matter are assumed to be coupled, so the radiation can be described by the local atmospheric conditions; radiation is said to be in local thermodynamical equilibrium (LTE). The same assumptions were made for the spectral synthesis carried out with the {\sc{bsyn}} code from the Uppsala package. 

The validity of these assumptions is a subject of discussion. Atmospheric convective motions can produce asymmetries in spectral line profiles. These asymmetries cannot be reproduced with the use of 1D model atmospheres, so in principle our results for isotopic abundance ratios based on observations of the profile asymmetry may be in error. A more appropriate treatment of the convection, like the one employed in the construction of hydrodynamic model atmospheres (3D), may be required for a better estimate of the line asymmetry due to the isotope lithium-6 \citep{Asplund06}. However Asplund et al. investigated the effects of the use of 3D model atmospheres on their $^6\mathrm{Li}/^7\mathrm{Li}$ calculations and found no significant differences with respect to the use of 1D models. The investigation was done assuming LTE conditions. Studies of the formation of the \ion{Li}{i} 670.7 line (e.g. \citealt{Carlsson94}) suggest that lithium abundances derived from this line observed in spectra of metal-poor stars may be affected by departures from LTE by up to 0.03\,dex. \cite{Cayrel07} showed that for the metal-poor star HD\,74000, 1D-LTE and 3D-NLTE results for $^6\mathrm{Li}/^7\mathrm{Li}$ disagree (0.024 vs 0.006). The 3D-NLTE  analysis of Cayrel et al. suggests there is no $^6\mathrm{Li}$ in the star. One of the purposes of our work is to compare our results with values reported in other works which are basically 1D-LTE results; that is why we have made a 1D-LTE assumption. The 3D-NLTE effects cannot be neglected when discussing implications of our results for the modelling of chemical evolution of the universe.
 
The \ion{Li}{i} 670.7 line is a multiplet and has two isotopic components, $^6\mathrm{Li}$ and $^7\mathrm{Li}$. The assumed values of wavelengths and $\log{gf}$ of these transitions are the same as in \cite{Smith98}. Laboratory measurements suggest they are known with a high accuracy. Atomic data for the \ion{Fe}{ii} lines used for metallicity determinations were taken from \cite{Nissen02} and from {\sc{vald}} \citep{VALD95} for lines other than these. 

\begin{table*}
%\begin{minipage}[t]{\columnwidth}
\caption{Observational information}
\label{stepar}
\centering
\renewcommand{\footnoterule}{}  % to avoid a line before footnotes
\begin{tabular}{lccccccccc}
\hline \hline
Star& $V$&Date & Exp. Time  & Inst. Setting $\lambda_0$ & $v_\mathrm{rad\ H}$&$T_{\mathrm{eff}}$ &$\log{g}$ & [Fe/H] & $\xi_t$ \\
 &(mag)&(UT)& (s) & (nm) &(km\,s$^{-1})$&(K)&(cgs)&(dex)&(km\,s$^{-1}$)\\
\hline

BD$-04^\circ\,3208$   & 9.99 & 18-05-05 & $4\times1800$ &620.78 (StdYb)& 58.7 & 6338   & 4.00   & $-2.28$  & 1.5 \\
 & &19-05-05 &$2\times1800$\\
G 64-37                       &11.14&  18-05-05 &$7\times1800$ &620.78 (StdYb) &82.1& 6318   & 4.16   & $-3.12$ & 1.5 \\
&&19-05-05 &$7\times1800$\\
BD$+02^\circ\,3375$  & 9.93&19-05-05 & $4\times1800$ &620.78 (StdYb) &$-397.1$&     5855   & 4.16   &$-2.19$  & 1.5 \\
BD$+20^\circ\,3603$  & 9.69 & 18-05-05 &  $1\times1800$ &620.78 (StdYb) &$-241.6$&   6092   & 4.04   & $-2.22$  & 1.5 \\
&&&$2\times2400$\\
&&19-05-05&$1\times1800$\\
BD$+26^\circ\,3578$  &9.37& 17-05-05  & $3\times1800$  & 608.11 (StdYd) &$-128.2$ &  6239   & 3.87   & $-2.33$  & 1.5 \\
&&&$1\times2400$\\
%081543  &     6514   & 4.23   & -2.72  & 1.5 & 17,18-05-05 & 620.78\\
\hline
\end{tabular}
%\end{minipage}
\end{table*}

\section{\label{secbroad} Stellar and instrumental broadening}

Our estimates for the isotopic abundance ratios ($^6\mathrm{Li}/^7\mathrm{Li}$) are based on the comparison of the observed and synthesised profile of the resonance \ion{Li}{i} 670.78~nm doublet line. Before the comparison, the synthetic spectrum has to be convolved by a function that reproduces the instrumental and the stellar line broadening; for this work we have assumed a Gaussian function. We adopted the approach of including the stellar and instrumental broadening contribution in a single free parameter, the broadening parameter $\zeta_\mathrm{broad}$ ($= \Delta\lambda_\mathrm{FWHM}$ [km\,s$^{-1}$]), that was determined from the analysis of up to 26 observed spectral lines:  \ion{Na}{i} (588.9\,nm), \ion{Mg}{i} (470.2, 552.8\,nm), \ion{Ca}{i} (558.8, 612.2, 616.2, 643.9\,nm), \ion{Ti}{ii} (444.3, 446.8, 450.1, 453.4, 456.3, 457.1\,nm), \ion{Fe}{i} (441.5, 452.8, 491.8, 492.0, 523.3, 526.9, 532.8, 537.1, 539.7, 561.5\,nm) and \ion{Fe}{ii} (458.3, 492.3, 501.8\,nm). 

Different $R$ values for the different lines could compromise the accuracy of our determinations, however $R$ measurements from ThAr lamp lines in the range $\lambda = 550$ to 700\,nm did not show significant systematic variations with wavelength ($\Delta R \sim 1\%$). The spread of values at a given wavelength was higher, though still not so significant ($\sim \pm 5\%$).  

The other two free parameters of the analysis were the rest wavelength of the observed spectra and the chemical abundance ($\log\epsilon_X$) of the line studied \footnote{$\log\epsilon_X = \log{N_{X}/N_\mathrm{H}} + 12$}. The free parameters were varied until the best fit to the observations was achieved, using the $\chi^2$ technique. 

\begin{figure}
   \centering
   \includegraphics[angle=0,width=9cm]{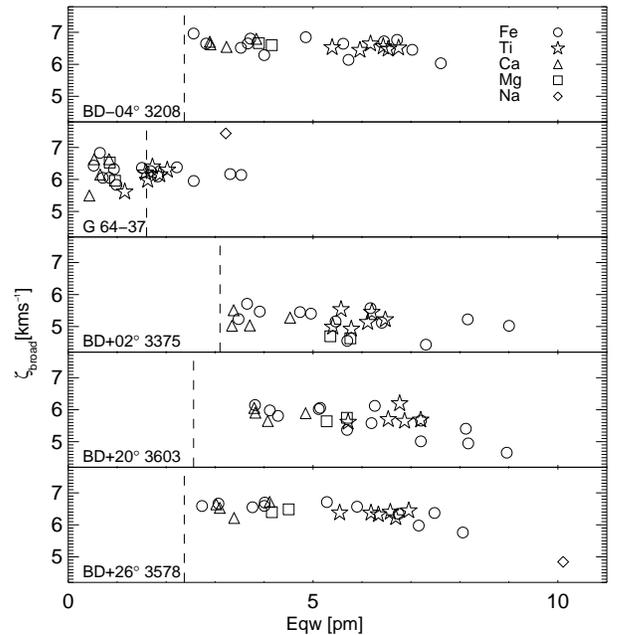}
      \caption{Broadening parameter versus the measured equivalent width of the different spectral lines used in the calculations. Lines of different atoms are shown with different symbols: Fe (circles), Ti (stars), Ca (triangles), Mg (squares) and Na (diamonds).}
         \label{macro}
   \end{figure}

Values of the broadening parameter versus the measured equivalent widths of the spectral lines are shown in Fig.\,\ref{macro} for all the observed stars of our sample. Not all lines were measurable in all stars. The data in Fig.\,\ref{macro} suggest that lines stronger than $\sim 7$\,pm require lower broadening values so mean values from the weaker lines were determined for $\zeta_\mathrm{broad}$ and taken as our best estimate, see Table\,\ref{broadpar}. The weakest lines observed have equivalent widths in the range 0.4 to 3.8\,pm depending on the star. For comparison, the observed Li lines have equivalent widths in the range 1.6 to 3.1\,pm (see Table\,\ref{lithium}), values are shown in Fig.\,\ref{macro} as dashed vertical lines; only  \object{BD $+20^\circ\,3603$} has a lithium line that is much weaker than the minimum equivalent width included in its broadening calculations (2.6 vs 3.8\,pm). We assumed that for this star the broadening parameter does not change significantly with equivalent width in the weak-line regime. 

\begin{figure}
   \centering
   \includegraphics[angle=0,width=9cm]{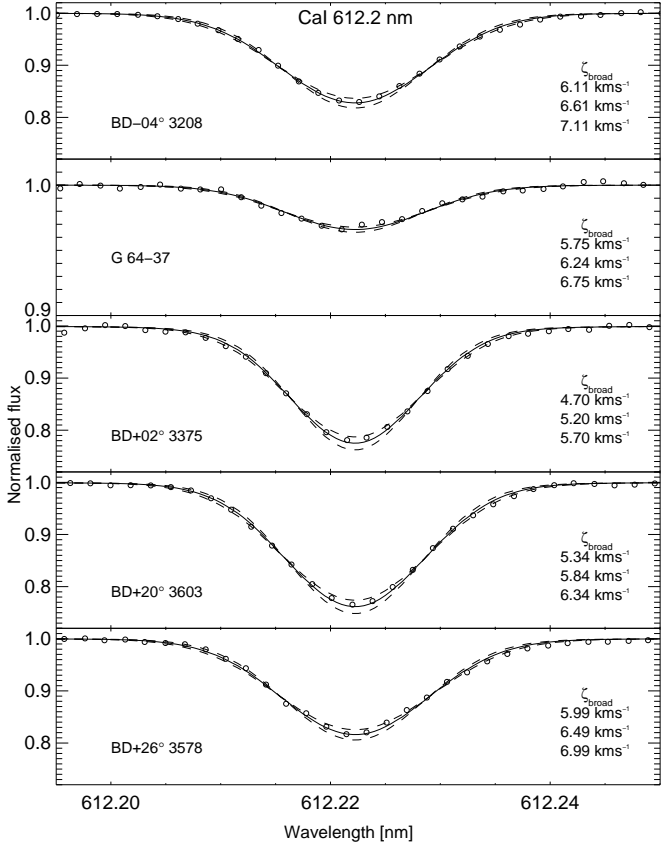}
      \caption{Spectral observations of the \ion{Ca}{i} 612.2 line for the five observed metal-poor stars (circles) and spectral synthesis for three values of the broadening parameter, the best estimate (solid line) and the best estimate $\pm$ 0.5 km/s (dashed lines).}
         \label{cai}
   \end{figure}

The mean values of the broadening parameter, listed in Table\,\ref{broadpar}, were well determined for all the stars with typical standard deviation values of 0.15-0.40\,km\,s$^{-1}$. The observed \ion{Ca}{i} 612.2 line and its best fit is shown in Figure\,\ref{cai} where you can appreciate the low sensitivity of the fit to changes in the broadening parameter of the order of $\pm0.5\,\mathrm{km}\,\mathrm{s}^{-1}$, this value is higher than the maximum of the $\sigma_\zeta$ values listed in column 3 in Table\,\ref{broadpar}.
%_____________________________________________________________
%                                    One column rotated figure
%-------------------------------------------------------------
%

%_____________________________________________________________
%                                    One column rotated figure
%-------------------------------------------------------------

\section{\label{seciso} Isotopic abundance ratios}

\subsection{Calculations}

The $^6\mathrm{Li}$ isotope is shifted by $\sim 0.015$\,nm relative to $^7\mathrm{Li}$ in the neutral lithium resonance doublet line. This produces an asymmetry in the already asymmetric profile of the red wing of the line. Under the assumptions of 1D model atmospheres, observed asymmetries may only be due to the multiplet configuration of the lithium resonance line and to the isotopic shifts (for 3D effects, see discussion in \S\,\ref{secstepar}). 

The best fit to the observations of the lithium line was determined from a $\chi^2$ analysis ($\chi^2 = \sum{(O_{i}-S_{i})^2/\sigma^{2}_\mathrm{S/N}}$; where $O_{i} = $ normalised observation in pixel $i$, $S_{i} =$ normalised synthesis and $\sigma_\mathrm{S/N} =$ noise based on the $S/N$) with three free parameters. A first, approximate, continuum normalisation was done in {\sc{iraf}}. The continuum level of the spectra was redetermined and fixed prior to the $\chi^2$ analysis, not taken as a free parameter as in \cite{Asplund06}. The continuum levels were determined by linear interpolation between two windows (see Figure\,\ref{tell}), one on each side of the lithium line. Different regions were selected for different stars in order to minimise the presence of structure.  The presence of ripples that remain after this continuum normalisation compromises the accuracy of the continuum level as well as the determination of the $S/N$. We return to this issue below. 

\begin{table}
\begin{minipage}[t]{\columnwidth}
\caption{Broadening parameter values and their standard deviation.}
\label{broadpar}
\centering
\renewcommand{\footnoterule}{}  % to avoid a line before footnotes
\begin{tabular}{lccccc}
\hline \hline
Star & $\zeta_\mathrm{broad}$ & $\sigma_\zeta$ \\
&(km\,s$^{-1})$&(km\,s$^{-1}$)\\
\hline
BD $-04^\circ\,3208$  &    6.61 &    0.18 \\
G 64-37    &  6.24    & 0.39 \\ 
BD $+02^\circ\,3375$   &   5.20   &  0.32 \\
BD $+20^\circ\,3603$    &  5.84    & 0.23\\
BD $+26^\circ\,3578$    & 6.49    & 0.16 \\

\hline
\end{tabular}
\end{minipage}
\end{table}

\begin{table*}
%\begin{minipage}[t]{\columnwidth}
\caption{Equivalent widths and $S/N$ measurements of the \ion{Li}{i} 670.78\,nm together with the best results for the $\chi^2$ three free parameters and the sensitivity of the analysis.}
\label{lithium}
\centering
\renewcommand{\footnoterule}{}  % to avoid a line before footnotes
\begin{tabular}{lccccccccccc}
\hline \hline
Star & $W_\lambda$& $S/N$ &$ \Delta\lambda$ & $A(^7$Li$)$ &\multicolumn{3}{c}{\isor}&&\multicolumn{3}{c} {$\Delta^6\mathrm{Li}/^7\mathrm{Li}$} \\
\cline{6-8}
\cline{10-12}

%\cline{8-11}
%\rule{0pt}{2.6ex}
&(pm)& &(nm)&(dex)\\

&&&&&ncor &cor & best-esti && cont   &  $\Delta\lambda_\mathrm{cover}$& $\zeta_\mathrm{broad}-\sigma_\zeta$\\
\hline
BD $-04^\circ\,3208$ &  2.37& 602 & 0.01036& 2.27 & 0.047& $0.043$ &$0.047\pm0.039$&&  $-0.037$ & 0.003& 0.009\\
G 64-37 & 1.60 & 496 & $-0.00125$& 2.04  & 0.006 &$0.000$ &$0.006\pm0.039$&& 0.028&0.023&0.013\\
BD $+02^\circ\,3375$ &  3.10 & 417& 0.00745 & 2.04 & 0.007 & $-0.001$ &$0.007\pm0.029$&&0.023 &0.007& 0.013 \\
%BD$+20^\circ\,3603$ &  2.56&492 & 0.00713 & 2.14 & $4.026\pm3.29$& 0.0079 & 0.0261 & $-0.0028$ & 0.0114&$-0.0011$&0.0019\\
BD $+20^\circ\,3603$ &  2.56&437& 0.00626& 2.14 & 0.004&0.025 & $0.025\pm0.025$ & &  $-0.003$ &0.002& 0.011\\

BD $+26^\circ\,3578$ &  2.37& 611&  0.00163 & 2.21 &0.074 &$0.004$ &$0.004\pm0.028$&&  0.017 &  0.014& 0.016\\

\hline
\end{tabular}
%\end{minipage}
\end{table*}

\begin{figure}
   \centering
   \includegraphics[angle=0,width=9cm]{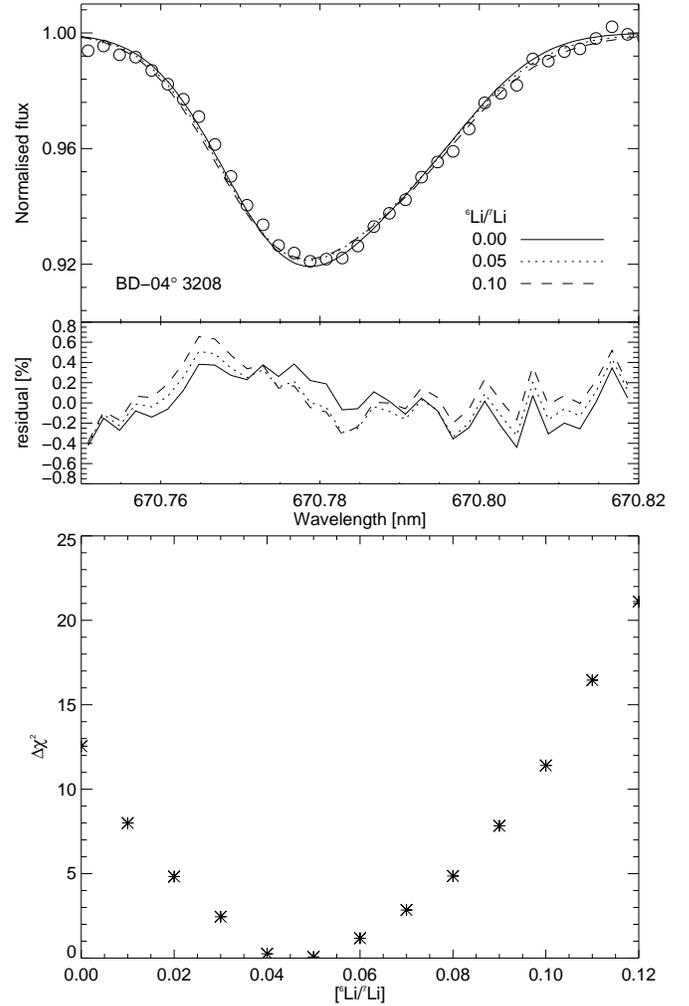}
      \caption{Top: the observed spectrum of the \ion{Li}{i} 670.78\,nm line for \object{BD $-04^\circ\,3208$} (circles) along with synthetic spectra for three different $^6\mathrm{Li}/ ^7\mathrm{Li}$ values:  0.00 (solid), 0.05 (dotted) and 0.10 (dashed). Middle: differences between observations and spectral synthesis given in $\%$. Bottom: non-reduced $\chi^2$ (three free parameters, 36 wavelength points) variations with {\isor}; the values are given relative to the minimum ($\Delta{\chi}^2$).}
         \label{cai1}
   \end{figure}

\begin{figure}
   \centering
   \includegraphics[angle=0,width=9cm]{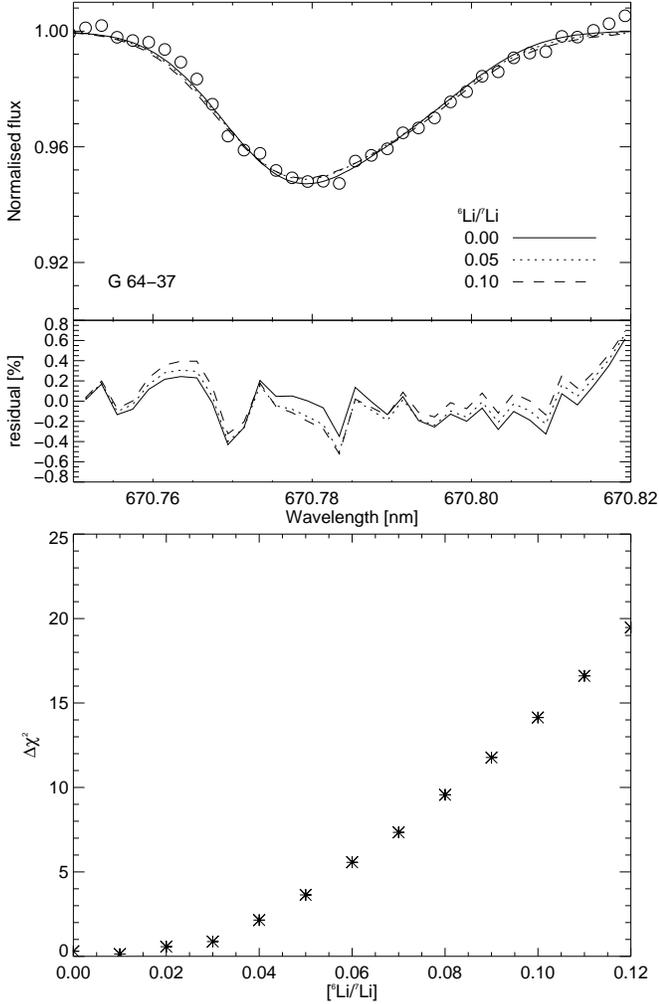}
      \caption{Same as in Figure\,\ref{cai1} but for \object{G 64-37} and 37 wavelength points.}
         \label{cai2}
   \end{figure}

\begin{figure}
   \centering
   \includegraphics[angle=0,width=9cm]{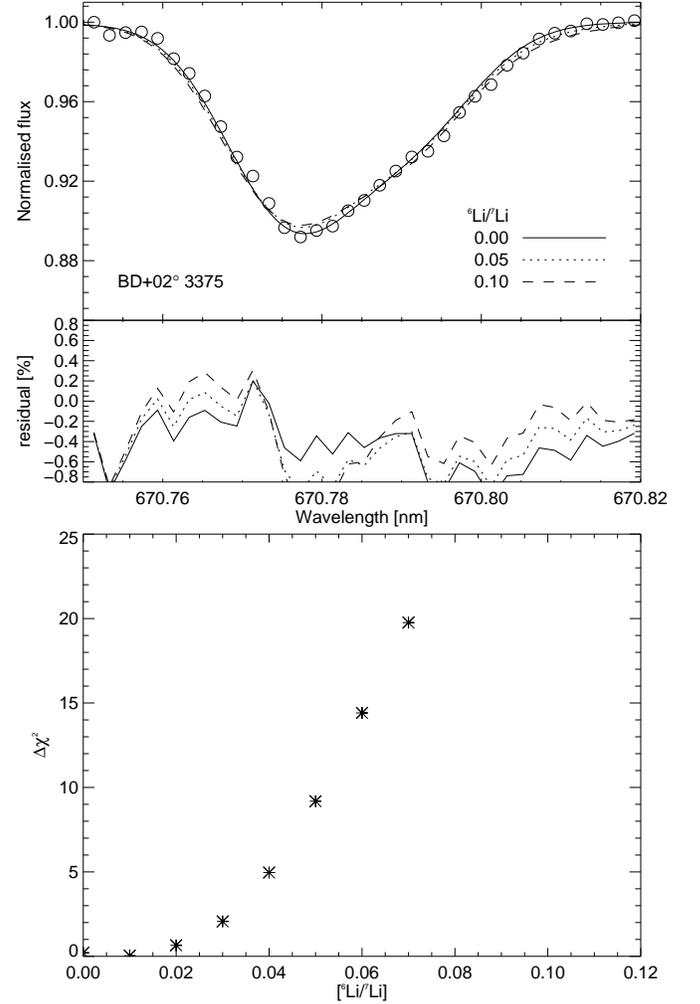}
      \caption{Same as in Figure\,\ref{cai2} but for \object{BD $+02^\circ\,3375$}.}
         \label{cai3}
   \end{figure}

With the broadening parameter and continuum level fixed, the free parameters of the $\chi^2$ analysis were the lithium-7 abundance $A(^7\mathrm{Li})$, the isotopic lithium abundance ratio $^6\mathrm{Li}/^7\mathrm{Li}$ and the wavelength zero point of the observations.The three free parameters were allowed to vary until $\chi^2$ was minimised. They were changed in steps of: 0.02~dex in lithium abundance, 0.01 in isotopic abundance ratios and 0.0002\,nm in wavelength shift, with typically a $11\times13\times11$ points grid. Wavelength points in a range from 670.743 to 670.823\,nm were included in the $\chi^2$ calculations (36-37 wavelength points). Our best {\isor} estimates for all the stars are listed in Table\,\ref{lithium} (column 8) together with the $S/N$ around the resonance neutral lithium line (column 3), the wavelength shift (column 4)  and the lithium-7 abundance (column 5). Figures\,\ref{cai1}, \,\ref{cai2}, \,\ref{cai3}, \,\ref{cai4} and \,\ref{cai5} show the observed metal-poor spectra and their best synthetic fits for {\isor}\ = 0.00, 0.05, 0.10 together with their residuals (observations $-$ synthesis in $\%$). The goodness of the final fits can be appreciated from the $\Delta\chi^2$ values ($\Delta\chi^2 = \chi{^2}-$ the minimum value) presented in these figures. Note that the $\chi{^2}$ values were not divided by the degree of freedom so the values are non-reduced $\chi{^2}$ values.

\subsection{\label{contam} Contamination}

%-------------------------------------------------------------
   \begin{figure*}
   \centering
   \includegraphics[angle=0]{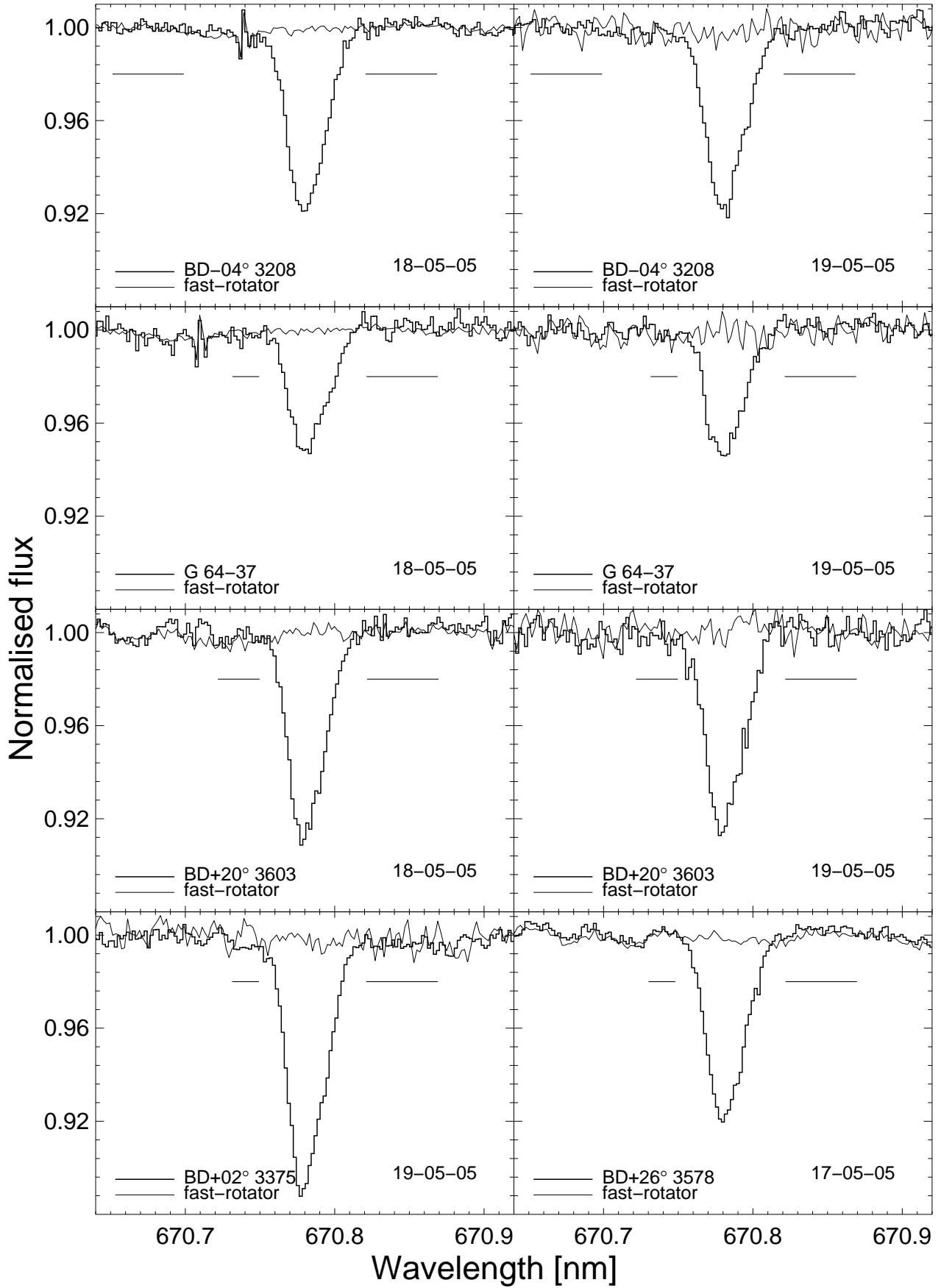}
      \caption{Observations around the \ion{Li}{i} 670.78\,nm line of the programme stars (thick histogram) and the associated fast rotator stars (thin line).}
         \label{tell}
   \end{figure*}

Possible ``contamination'' of the lithium line was investigated by comparing the spectra of the observed metal-poor stars with the spectra of hot fast rotator stars taken the same night. This comparison is shown in Figure\,\ref{tell} where two plots are shown for stars observed in two different nights, each corresponding to a different date of observation. The associated hot fast rotator spectra of the four metal-poor stars \object{G 64-37}, \object{BD $+02^\circ\,3375$}, \object{BD $+20^\circ\,3603$} and \object{BD $+26^\circ\,3578$} show tiny features of peak-to-valley range $\sim 0.003$-$0.006$ (in normalised flux units) that suggest that their observed lithium line suffer from some contamination. At most one or two of such features could be identified with diffuse interstellar bands. We presume the majority is residual fringing or residual flat-field errors. Note that, although the observed ripples would not significantly affect common abundance determinations, they can have a great impact on the isotopic abundance ratios. One approach to quantify the effect of these ripples in the {\isor} results is to divide the metal-poor spectra by the spectra of the hot fast rotator stars. The division has two effects: one is to alter the flux in the Li line directly, and the other is to require a revision of the continuum fit, which again, affects the normalised flux in the line. The correction is not perfect; there are features in the observed metal-poor spectra that do not match the features in the fast-rotator spectra (e.g. \object{BD $-04^\circ\,3208$}) so we cannot be sure the lithium line is free of ripples at the level of accuracy that we require for our determinations even after this correction. Results for the isotopic abundance ratios coming from non-corrected and the corrected (divided) spectra are presented in Table\,\ref{lithium}, columns 6 and 7 respectively.

The observations of \object{BD $+02^\circ\,3375$} were from only one night and, unfortunately, the $S/N$ of the fast rotator spectra for that night was not very high, $S/N \sim 250$. To avoid degrading the quality of the analysed spectra, the division by a fast rotator spectrum should be done only when necessary and so for this star it was not done. Had the division been done for this star, the signal-to-noise of the analysed observations would have decreased ($S/N \sim437$ to 300) but the {\isor} results would have not changed significantly, 0.007 vs $-0.001$ (see values in Table\,\ref{lithium}). For this star we have adopted the {\isor} value based on the non-corrected spectra (higher $S/N$) as our best estimate, while the ripple effect ({\isor}$|_{cor}-$\isor$|_{ncor}$) is included in the error bar. 

The problem of the low $S/N$ of the fast rotator spectrum was not such a big issue for the other stars observed that night,  \object{G 64-37}, \object{BD $-04^\circ\,3208$} and \object{BD $+20^\circ\,3603$}, as they were observed also on a second night. Therefore, the $S/N$ per night required to achieve the high target $S/N$ was lower than for the stars which were observed on only one night. Still, the low $S/N$ makes it difficult to judge by eye whether there is, or not, contamination. 

Observations of  \object{G 64-37} suggest no contamination for the first night, whereas it is hard to say for the second night. The {\isor} corrections for \object{G 64-37} are very small, $\Delta${\isor\ }$\sim 0.007$, so for this star as well, no correction was done. The comparison of the spectra of \object{BD $-04^\circ\,3208$} and its associated fast rotator does not show any sign of strong contamination. Hence, our isotopic ratio results coming from non-corrected and corrected spectra are quite similar, 0.047 vs 0.043, our best estimate is based on the non-corrected spectra.

Results for \object{BD $+20^\circ\,3603$} were more affected by the division of spectra: {\isor\ }$\sim 0.004$ (non-corrected) vs $0.025$ (corrected). We consider it more appropriate for our best estimate to adopt a value based on the corrected spectra, given that there are indications of contamination. 

\begin{figure}
   \centering
   \includegraphics[angle=0,width=9cm]{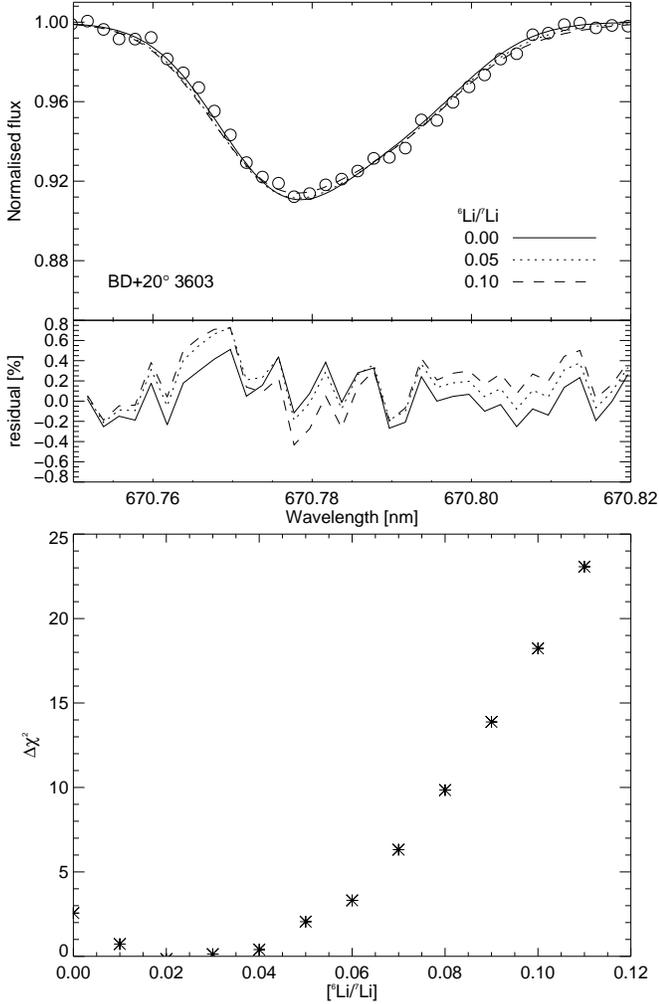}
      \caption{Same as in Figure\,\ref{cai2} but for \object{BD $+20^\circ\,3603$}. }
         \label{cai4}
   \end{figure}

\begin{figure}
   \centering
   \includegraphics[angle=0,width=9cm]{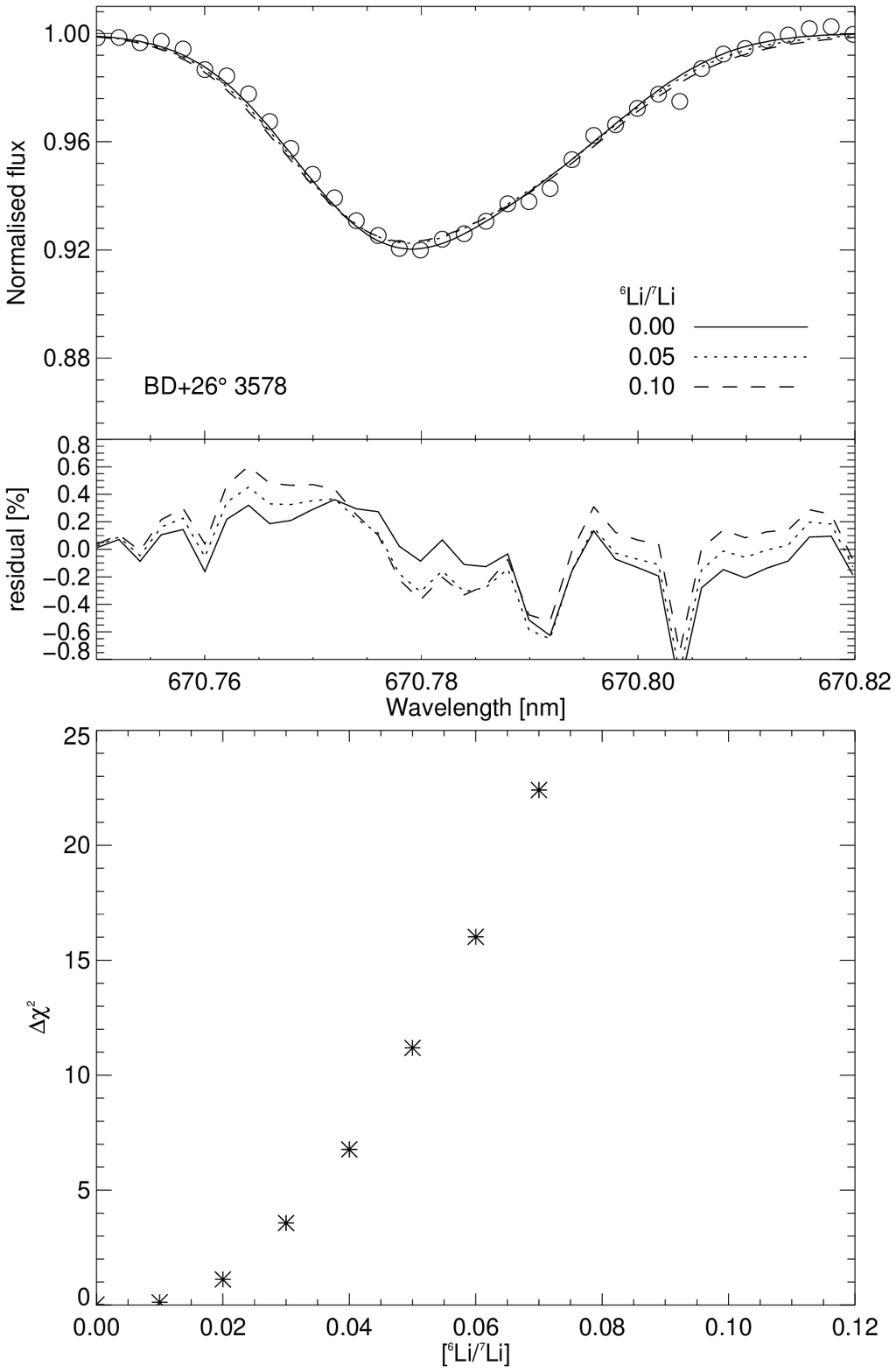}
      \caption{Same as in Figure\,\ref{cai2} but for \object{BD $+26^\circ\,3578$}. }
         \label{cai5}
   \end{figure}

This was also the case for \object{BD $+26^\circ\,3578$}; ripples are observed in the hot spectrum at the wavelengths of the \ion{Li}{i} 670.8\,nm line, therefore we applied the correction to get our {\isor} best estimate, {$0.004$} (corrected) vs  {$0.074$} (non-corrected). The match between the ripples in the metal-poor spectrum and in the associated fast rotator is quite impressive (see Figure\,\ref{tell}). The observed residual fringing in the fast rotator spectra recorded on May $17^{\mathrm{th}}$, 2005 was greater than on the other nights when a different instrumental setting was used.  \object{BD $+26^\circ\,3578$} was the only programme star observed on that night (see Table\,\ref{stepar}). The observed spectra shown in Figures\,\ref{cai4} and \ref{cai5} are the divided spectra.  

\subsection{\label{sensi} Sensitivity}

The sensitivities of our results to the different assumptions made in the analysis have been studied. Concerning the normalisation, changes of $0.1\%$ in the continuum level, where the typical line depth is $10\%$, has an effect of 0.02 on the {\isor} results, so the normalisation of the spectra is an important issue. 

After the initial continuum fit (see $\S$\,\ref{obs}), the spectrum of \object{BD $+26^\circ\,3578$} showed a remaining tilt, so it was re-normalised in {\sc{iraf}} using a linear function similar to that employed in the $\chi^2$ analysis. After this was done, the match between the ripples in the spectrum of \object{BD $+26^\circ\,3578$} and in the associated fast rotator spectrum was excellent, hence we have used a linear fit in our analysis. For stars with only minor differences in the measured continuum level on either side of the lithium line, the use of the mean value of the continuum fluxes, rather than a linear interpolation, does not have a big effect on the {\isor} results. However, for \object{BD $+26^\circ\,3578$} and \object{BD $+02^\circ\,23375$}, the isotope ratio changed by $\Delta${\isor\ }$= 0.005$ when using the mean-value approach. 

The level measurements can change with the spectral window choice, different areas were selected for different stars and they are shown by horizontal bars in Fig.\,\ref{tell}. A change of choice within the range of possibilities would affect the results significantly for all the stars except \object{BD $+20^\circ\,3603$}. Even a shift of the window in the left wing by $<0.1$\,nm will change the isotope abundance ratios significantly, the changes ranging from $\Delta${\isor\ }$= -0.037$ (\object{BD $-04^\circ\,3208$}) to $0.028$ (\object{BD $+02^\circ\,3375$}), see column 9 in Table\,\ref{lithium}. \object{BD $-04^\circ\,3208$} and \object{BD $+02^\circ\,3375$} are the stars with the highest uncertainty in the continuum location among our sample. For these two stars, the use of the fast rotator spectrum did not help for discriminating between two different windows in the left wing, either because the small structures observed in the fast rotator spectra did not match the ones in the metal-poor spectra or because the spectra of the comparison star had a low $S/N$. The result of this was that two normalisation levels were possible. Summarising, isotopic results for these stars have a high uncertainty because of the uncertain continuum location.
  
Another aspect to take into account is the number of wavelength points included in the $\chi^2$ calculation, in particular in the red wings of the lithium line. A decrease of typically four or five wavelength points ($\sim 0.010$\,nm) in the red wing affects the isotope ratio results by a few tenths of one percent for three stars (Table\,\ref{lithium}, column 10). The stars \object{G 64-37} and \object{BD $+26^\circ\,3578$} were more affected by these changes, $\Delta${\isor\ }= 0.023 and 0.014 respectively. 

Results also show sensitivities to the assumed values of macroturbulence; we determined new isotope ratio results for broadening parameters values decreased by their $\sigma_\zeta$. We find a mean sensitivity of $\Delta${\isor\ }$\sim 0.01$ (see values in column 11 of Table\,\ref{lithium}), although the sensitivity for \object{BD $+26^\circ\,3578$} is even higher, 0.016.

We have investigated the effects of changing the stellar parameters for one particular case, \object{BD $-04^\circ\,3208$}, according to their uncertainty: $\Delta T_\mathrm{eff} = 100\,$K, $\Delta\log{g} = 0.2$, $ \Delta[\mathrm{Fe}/\mathrm{H}] = 0.2\,$dex and $\Delta \xi_\mathrm{t} = 0.2\,$km\,s$^{-1}$. Isotopic ratio determinations are not very sensitive to these changes directly, $\Delta^6\mathrm{Li}/^7\mathrm{Li}<$ 0.004 (see top row of Table\,\ref{senstepar}). A second effect of changing the stellar parameters is to alter the inferred broadening parameter (see second row of Table\,\ref{senstepar}). For example, an increase in the microturbulence results in lower $\zeta_\mathrm{broad}$ values; the broadening parameter compensates particularly for those parameters most strongly connected with the profile width: $\Delta \xi_\mathrm{t}$ and $\log{g}$. We have shown already (Table\,\ref{lithium}, column 11) that the isotopic ratio is quite sensitive to the broadening parameter. The total effect of increasing the microturbulence will be a {\isor} increase of $\sim 0.003$ (fourth row of Table\,\ref{senstepar}), $-0.003$ from microturbulence directly, and $+0.006$ from the resulting change in $\zeta_\mathrm{broad}$ (third row of Table\,\ref{senstepar}). A total error associated with the uncertainty in the stellar parameters including the broadening parameter is estimated from the quadrature sum of the values in fourth row in Table\,\ref{senstepar} and is $\Delta${\isor\ }$\sim 0.006$. Note that this excludes the direct effect of the uncertainty in the broadening parameter, which is given in Table\,\ref{lithium}, column 11.  Summarising, the two first rows of Table\,\ref{senstepar} list the partial sensitivities to the stellar parameters, the third row the partial sensitivity of {\isor} results associated to the change in the broadening parameter as a result of changing the stellar parameters, and the last row the total combined effect (stellar parameters + broadening parameter). The quadrature sum of all the uncertainties are listed in the last column.

A summary of all the individual sensitivities are given as $\Delta${\isor} values in Table\,\ref{lithium} (columns 9 to 11) and the final row of Table\,\ref{senstepar}. The quadrature sum of these plus the {\isor}$|_{cor}-$\isor$|_{ncor}$ values were taken as our best estimates of the total {\isor} errors (Table\,\ref{lithium}, column 8). We considered the division by a hot fast rotator spectrum to be essential, not merely optional, for \object{BD $+26^\circ\,3578$}, so the {\isor} error bar for this star does not include the uncertainty associated with that correction, otherwise the error bar would be overestimated. 

Another concern is whether asymmetries are intrinsic to even single lines. Following \cite{Cayrel07}, we have studied the asymmetry of the observed profiles of two \ion{Fe}{i} lines at 639.36\,nm and 649.50~nm in the spectrum of \object{BD $-04^\circ\,3208$}. For that, the profiles were re-scaled to the same line depth, they were averaged and after that they were convolved with a Gaussian function of width 2.39 km\,s$^{-1}$ to correct for the different thermal broadening of the Fe and Li lines (see \citealt{Cayrel07}). The magnitude of the asymmetries shown in the determined profile depends on the assumed normalisation; the presence of fringing can introduce asymmetries. Continuum levels that are acceptable choices can produce an asymmetry whose additional depth is of the same order of magnitude as that produced by a displaced line of equivalent width $4\%$ of the main one. Therefore, despite the quality of our observations, it is difficult to determine whether the asymmetry is real or just a product of the normalisation. 

\begin{table*}
%\begin{minipage}[t]{\columnwidth}
\caption{Sensitivity of the {\isor} and the broadening parameter results to the stellar parameters uncertainties for \object{BD $-04^\circ\,3208$.}}
\label{senstepar}
\centering
\renewcommand{\footnoterule}{}  % to avoid a line before footnotes
\begin{tabular}{lcccccc}
\hline \hline
Star      &  $\Delta T_{\mathrm{eff}}$ &$\Delta\log{g}$ &$\Delta$ [Fe/H] & $\Delta \xi_t$ & $\Delta \zeta_\mathrm{broad}$ & $\sigma _ {^6\mathrm{Li}/^7\mathrm{Li}}$\\
&(K)&(cgs)&(dex)&(\,km\,s$^{-1}$)&(\,km\,s$^{-1}$)\\
& $+100$& $+0.2$& $+0.2$& $+0.2$ & $+0.2$\\
\hline
$\Delta^6\mathrm{Li}/^7\mathrm{Li}_\mathrm{direct}$ & $-0.004$ & $-0.004$ & $0.001$ & $-0.003$ & $-0.008$\\
$\Delta \zeta_\mathrm{broad}$ \footnote{km\,s$^{-1}$} & 0.03 & $-0.08$ & $-0.02$ & $-0.15$& &\\
$\Delta^6\mathrm{Li}/^7\mathrm{Li}_\mathrm{broad}$ & $-0.001$ & $0.003$ & $0.001$ & $0.006$ & & \\
$\Delta^6\mathrm{Li}/^7\mathrm{Li}_\mathrm{total}$ & $-0.005$ & $-0.001$ & $0.002$ & $0.003$ & & 0.006\\
\hline
$^3$ (\,km\,s$^{-1}$)\\
\end{tabular}
\end{table*}

\subsection{\label{summary}Summary}

Our best estimates of the {\isor} ratios in these stars are set out in Table\,\ref{lithium}, column 8 and displayed in Fig.\,\ref{asplund}. The tabulated uncertainty is a quadrature sum of the contribution from uncertainties in: whether or not division by a hot fast rotator is performed (\S\,\ref{contam} and Table\,\ref{lithium}, columns 6 and 7), $\sim 0.010$, changes resulting from different choices of continuum windows (\S\,\ref{sensi} and Table\,\ref{lithium}, column 9), $\sim 0.022$; changes resulting from different choices of the wavelength range over which the observed and synthetic spectra are compared via the $\chi^2$ statistics (Table\,\ref{lithium}, column 10), $\sim 0.010$; the star's broadening parameter (Table\,\ref{broadpar}, column 3 and Table\,\ref{lithium}, column 11), $\sim 0.012$; and the star's atmospheric parameters (Table\,\ref{senstepar}, row 1), $\sim 0.006$.  We note, but do not include, further possible error sources: whether a mean value of the continuum windows or a linear interpolation between continuum windows is adopted as ``the continuum''; and whether even single spectral lines are asymmetric. The tabulated uncertainties range from $\pm 0.025$ to $\pm 0.039$, and are dominated by the uncertainty in the continuum, even though the $S/N$ exceeds 400 in every case. The reason that the continuum placement continues to exert such an influence on the outcome of the analysis lies substantially in the problem of small structures seen in the reduced data, and that although they are small by most subjective measures, they are significant relative to the very weak signal of $^6$Li, the spectral line of which has a central depth of probably only $\leq 0.003$ normalised flux units (i.e. $\leq 4\%$ of the depth of the $^7$Li, which in turn is $\sim 0.08$ flux units). Based on the extreme sensitivity of the isotope ratio results to alternative choices in the spectral analysis, we are certain that we could not infer {\isor} to better than $\pm0.025$, or perhaps more realistically $\pm 0.040$, despite having data of $R = 95\,000$ and $S/N \approx 400$ to 600. Moreover, we note that the formal errors which we quote are not the maximum possible values and in some cases will underestimate the real error, particulary since they result from the quadrature sum of several distinct contributions. The continuum from $\sigma_\zeta$ in particular is only a $1\sigma$ value.

\section{\label{secdis} Discussion}

 According to models of big bang nucleosynthesis and cosmic-ray spallation, low lithium-6 abundances are predicted in stars of low metallicity. Detections of this isotope in these stars would have thus a strong impact on production models of light elements. All the tests we have done, e.g. comparing the results based on non-corrected and corrected spectra, on different normalisation levels, on different wavelength ranges covered in the analysis, suggest that our best estimates for {\isor} are uncertain to at least 0.025-0.040. Given than our best estimates of {\isor} are of the same order of magnitude as these values, we do not claim any significant detection. 
 
\begin{figure}
   \centering
   \includegraphics[angle=0,width=9cm]{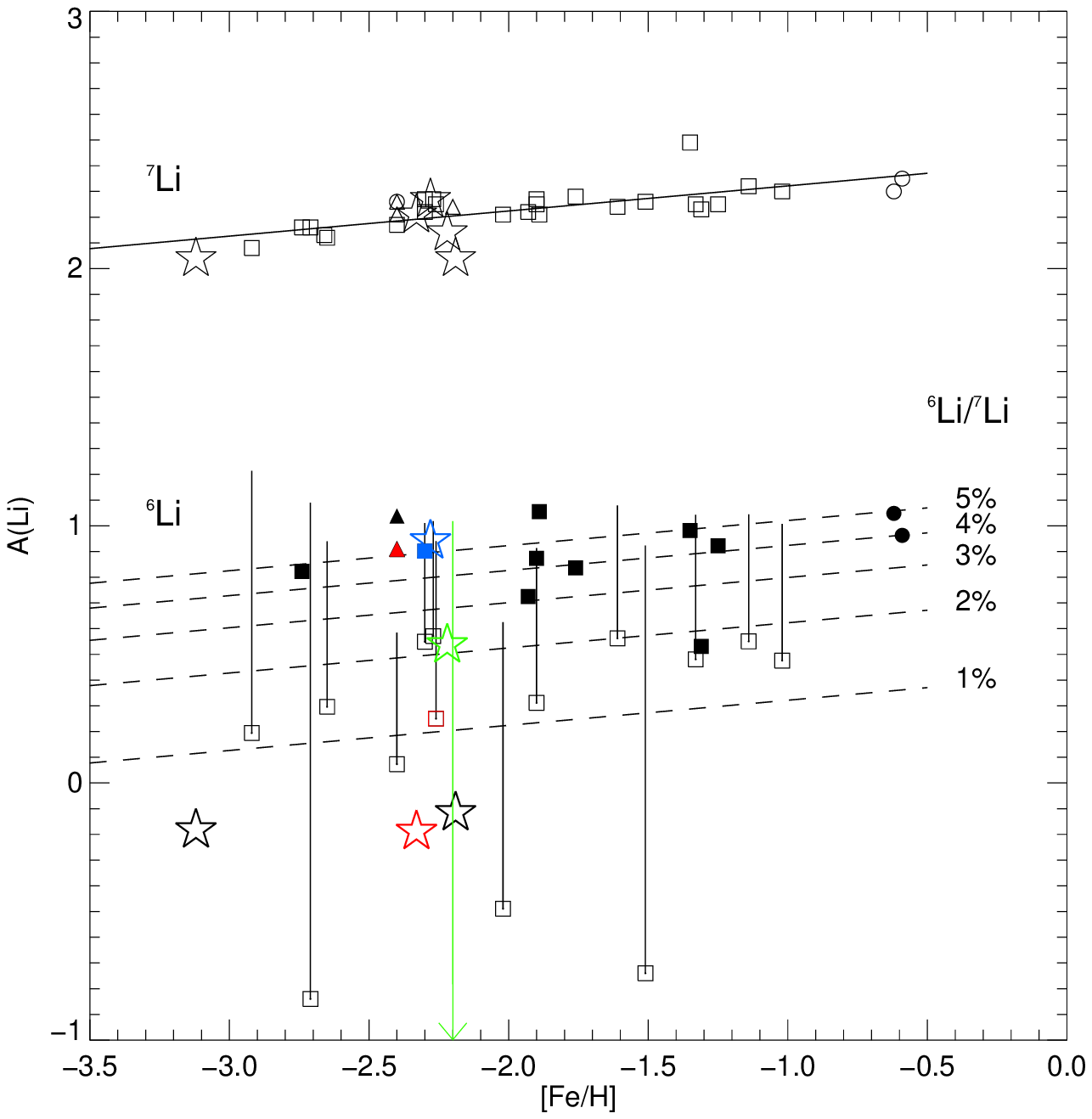}
      \caption{Derived logarithmic abundances of $^7\mathrm{Li}$ and $^6\mathrm{Li}$ versus stellar metallicities. The linear fit to $^7\mathrm{Li}$ abundances is presented (solid line) along with $^6\mathrm{Li}$ contributions (dashed lines). Our data are shown (stars) together with the LTE data from \citet{Asplund06} (squares), for the stars in common with our study also the data from \citet{Smith98} (triangles) and \citealt{Nissen99} (circles) are displayed. The filled symbols indicate detections, the bars are $3\,\sigma$ upper limits and the non-black colors denote stars in common. [{\it See the electronic edition of the Journal for a color version of the figure.}]}
         \label{asplund}
   \end{figure}

One of the goals of this paper was to observe targets whose lithium-6 content was already determined from observations taken with other spectrographs, and show whether, or not, the published results were corroborated by our observations. There are three objects in common with other studies:  \object{BD $+26^\circ\,3578$}, \object{BD $-04^\circ\,3208$} and \object{BD $+20^\circ\,3603$} (color symbols in Figure\,\ref{asplund}, {\it see the electronic edition of the Journal for a color version of the figure}). 

The first was observed in \citet{Smith98} and \citet{Asplund06} (red symbols in Fig.\,\ref{asplund}). They reported different isotopic ratios results:  $^6\mathrm{Li}/^7\mathrm{Li}=0.05\pm0.03$ ($S/N \sim 420$) and $^6\mathrm{Li}/^7\mathrm{Li}=0.01\pm0.013$ ($S/N \sim 680$) respectively. \citeauthor{Asplund06} suggested that the differences between their value and that of \citeauthor{Smith98} could be due to the differences in the determination of the broadening parameter. We have used a similar method to that employed by \citeauthor{Asplund06}, the difference being that we did not use a rotational profile (metal-poor stars have low rotational velocities). We obtain $\zeta_\mathrm{broad}=6.50$ vs 5.65\,$\mathrm{km}\, \mathrm{s}^{-1}$ \citep{Asplund06}. We have data of similar $S/N$ ($\sim 642$) but slightly lower resolving power ({$R \sim 95\,000\ \mathrm{vs}\ 115\,000$}). Our best estimate for this star ({\isor\ }$= 0.004\pm0.028$) agrees with the result from \citet{Asplund06}. Much higher values {\isor\ }$= 0.074$ would be determined from the non-corrected spectra, but based on the match between the residual features seen in our spectra of this star and the corresponding fast rotator (see Figure\,\ref{tell}, bottom right), we are confident that division (correction) by the fast rotator is essential for \object{BD $+26^\circ\,3578$}. Our observed bright fast rotator star spectra suggest a residual fringing of the order of $0.4-0.6\%$ (peak-to-valley) whereas \citeauthor{Asplund06} report a value much lower for theirs $\le 0.2\%$ ($\pm 0.1\%$). The amplitude of the fringe patterns of the HDS spectra (peak-to-valley difference) at 670.0~nm is typically $5\%$, which results in a pattern of a few tenth percent after flat-fielding. The width of the fringe pattern is much larger than the Li absorption feature, and the effect on the line analysis is usually not severe. We note, however, that the effect is not negligible in the setting adopted on May 17th 2005 (see \S 5.2). According to \citeauthor{Smith98}, their residual fringing is negligible although they do not quote any magnitude so it is difficult to judge whether fringing can account for the differences between the results. 

Our best estimate for \object{BD $-04^\circ\,3208$}, the other star in common with \citealt{Asplund06} (blue symbols in Fig.\,\ref{asplund}, {\it see the electronic edition of the Journal for a color version of the figure}) is $0.047\pm0.039$ which agrees well with their value $0.042\pm0.019$. Our error bar is higher so we do not claim any detection in contrast to them. Concerning the quality of the spectra, we have slightly higher $S/N$ ($\sim 600$ vs 530). 

The other star in common with \citet{Smith98} is \object{BD $+20^\circ\,3603$} (green symbols in Fig.\,\ref{asplund}). We have data of similar quality to theirs ($S/N \sim 430$), though their $R$ was a bit higher ($\sim115\,000$). Our best estimate of $0.025\pm0.025$ is higher than their best estimate of $0.00\pm0.02$ but consistent within the uncertainties. If no fast rotation correction was done to our spectrum, a {\isor} value closer to 0.00 would better reproduce the observations. Differences in the assumed stellar parameters for these stars are not significant so they should not affect the results.

Stellar parameters assumed for the stars \object{BD $-04^\circ\,3208$} and \object{BD $+26^\circ\,3578$} are similar to one another. The stars also have spectra of similiar quality. Their isotope abundance ratio results are different but agree within their error bars, $0.047\pm0.039$ vs. $0.004\pm0.028$; the continuum location for \object{BD $-04^\circ\,3208$} in particular is quite uncertain. A change in the continuum level within the uncertaintity for this star would bring the best {\isor} estimates for the two stars closer, $\sim 0.01$ vs $\sim 0.00$, though given that their temperatures differ by $\sim 100$\,K, they may have experienced different degrees of $^6\mathrm{Li}$ depletion \citep{Deliyannis90}. The stellar parameters values assumed for \object{BD $+20^\circ\,3603$} and \object{BD $+26^\circ\,3578$} are slightly different and as in the previous case, their isotopic estimates are different but agree within their error bars. The agreement gets better if no correction to the observed spectra is made for \object{BD $+20^\circ\,3603$}. According to our error estimates, the {\isor} ratios of these two stars are the most accurate.
  
We are not aware of any other lithium-6 determinations in the literature for the two targets \object{BD $+02^\circ\,3375$} and \object{G 64-37}. Unfortunately, the data quality for these two stars were among the lowest of the observed sample. Our best estimate suggests that we can only determine an upper limit of $\sim 0.03$ (1$\sigma$) for \object{BD $+02^\circ\,3375$}. The temperature assumed for this star is the lowest among our sample and low enough, according to the isochrones in {\cite{Deliyannis90}, for the star to have suffered from severe $^6\mathrm{Li}$ depletion ($\sim 95\%$ depletion). In fact, published detections of $^6\mathrm{Li}$ at metallicities similar to that of \object{BD $+02^\circ\,3375$ are for hotter stars. Concerning \object{G 64-37}, the most metal-poor star in our sample, we could estimate only an upper limit of {\isor\ }$\sim 0.04$ (1$\sigma$)}. \citealt{Asplund06} reported a $^6\mathrm{Li}$ detection for other star of $[\mathrm{Fe}/\mathrm{H}]<-2.5$, LP 815-43 ({\isor}$= 0.046\pm0.022$).}

The $^6\mathrm{Li}/^7\mathrm{Li}$ values do not correlate with $S/N$; we determined high isotopic ratio values from both high $S/N$ spectra (\object{BD $-04^\circ\,3208$}) and low $S/N$ (\object{BD $+20^\circ\,3603$}), and the same applies to low values of the isotopic ratio.

The principle conclusion of this study concerns not the isotope ratio itself but rather the uncertainties associated with the measurements. Despite having data of high resolving power ($R = 95\,000$) and $S/N \sim$ 400 to 600, a thorough examination of error contributions has led us to conclude that, for our dataset, we could not confidently determine the isotope fraction to better than $2.5\%$ to $4.0\%$. Moreover, the detections achieved by other groups for metal-poor stars are around this limit, as we show with the $^6\mathrm{Li}$ contribution functions in Figure\,\ref{asplund} determined from the linear fit to the $^7\mathrm{Li}$ observed abundances. It is quite possible that other studies have achieved a higher precision than was achieved in our study, despite the similarity of the initial datasets. Although, notice that \citealt{Asplund06} observations show lower levels of fringing than ours.   In contrast, we interpret our results as indicating that extreme caution is required in finally quantifying the uncertainties, in order to establish {\isor} ratios to the level of precision required to be confident of detections of $^6$Li at the $4\%$ level or lower. We are not confident that we have significant detections of $^6$Li in any of our five stars.

\section{\label{seccon} Conclusions}

The observed spectra of the \ion{Li}{i} resonance line were analysed for five metal-poor stars. Our best fits to the observations suggest that only the star \object{BD $-04^\circ\,3208$} may have a high lithium-6 content, {\isor}\ $= 0.047 \pm 0.039$. However, the uncertainties of the calculations are comparable to the calculations themselves, such that positive detections cannot be claimed for any of them. In general the analysis is very sensitive to the assumptions made for the continuum location, the residual fringing treatment and even the wavelength range covered in the analysis. Our error bars ($\Delta${\isor\ }$\sim 0.025$-$0.040$) tell us that detections may have been possible if the stars had a higher lithium-6 content  ({\isor\ }$\geq$ 0.050-0.080) following a $2\sigma$ criteria. 

Not surprising, we observed asymmetries in the profile of the \ion{Fe}{i} lines as \citet{Cayrel07} did. However, we have major concerns because the magnitude of the asymmetries depends on the spectral normalisation carried out, and the presence of residual fringing could also introduce spurious asymmetries. 

Our {\isor} results are comparable to recent values in the literature, but our error estimates are higher, the result of a detailed analysis of a range of uncertain aspects of the analysis associated with the significant presence of structure in our reduced data-set (summarised in \S\,\ref{summary}). In contrast, \citealt{Asplund06} tabulated only the uncertainties due to finite $S/N$ ---which they find usually to be their dominant error--- but not the wide range of error sources encountered in our work. This puts limitations on the detections so that only upper limits of {\isor} were inferred for the stars observed in this work.

One possible, but unjustified, generalisation of our results would be to question the reliability of all {\isor} measurements at levels of 4\% or 5\%. Making this generalisation would then call into question the evidence for a possible upper envelope or plateau in the $^6$Li abundance, which has caught the interest of astrophysicists in recent years. We believe there is a value in continuing research into the {\isor} ratios in metal-poor stars, including the fullest treatment of possible error sources, to determine whether such an upper envelope or plateau exists. 

\begin{acknowledgements}
      The first author is deeply grateful to Prof. Martin Asplund, Prof. Poul Erik Nissen and Prof. Bengt Gustafsson for their fruitful discussions and to Prof. Nikolai Piskunov for his assistance with data reduction. She is grateful to the UK \emph{Royal Society} for funding several research visits, one to the Astronomical Observatory of Japan and the others to the Uppsala University. This work has been supported by a research grant from the UK \emph{Science \& Technology Facilities Council.} 
\end{acknowledgements}

\bibliographystyle{aa}
 \bibliography{12289}
 % \end{thebibliography}
 \end{document}